\documentclass[12pt,letter]{article}

\usepackage{subfigure}
\usepackage{graphicx, epsfig, color}
\usepackage{amsmath,amssymb}
\def\bi{\begin{itemize}}
\def\ei{\end{itemize}}

\def\ta{\tilde a}

\def\sps1ap{SPS1a$^\prime$}
\def\c1p{C1$^\prime$}

\def\ts{\tilde s}

\def\tg{\tilde g}

\def\tq{\tilde q}

\def\be{\begin{equation}}  
\def\ee{\end{equation}}  
\def\bea{\begin{eqnarray}}  
\def\eea{\end{eqnarray}}  
\def\beas{\begin{eqnarray*}}  
\def\eeas{\end{eqnarray*}}  
\newcommand\prd[3]{{\it Phys.\ Rev.\ }{\bf D #1} (#2) #3}

\newcommand\prl[3]{{\it Phys.\ Rev.\ Lett.\ }{\bf #1} (#2) #3}
\newcommand\plb[3]{{\it Phys.\ Lett.\ }{\bf B #1} (#2) #3}
\newcommand\jhep[3]{{\it J. High Energy Phys.\ }{\bf #1} (#2) #3}

\newcommand\npb[3]{{\it Nucl.\ Phys.\ }{\bf B #1} (#2) #3}

\begin{document}
\begin{titlepage}
\begin{flushright}
FTPI-MINN-14/36
\end{flushright}

\vspace{0.5cm}
\begin{center}
%{\Large \bf Radiatively-driven natural Little Hierarchy
%}\\ 
{\Large \bf A natural Little Hierarchy for SUSY\\
from radiative breaking of PQ symmetry
}\\ 
\vspace{1.2cm} \renewcommand{\thefootnote}{\fnsymbol{footnote}}
{\large Kyu Jung Bae$^{1,2}$\footnote[1]{Email: bae@nhn.ou.edu },
Howard Baer$^{1,2}$\footnote[2]{Email: baer@nhn.ou.edu },
and Hasan Serce$^{1,2}$\footnote[3]{Email: serce@ou.edu } 
}\\ 
\vspace{1.2cm} \renewcommand{\thefootnote}{\arabic{footnote}}
{\it 
$^1$Dept. of Physics and Astronomy,
University of Oklahoma, Norman, OK 73019, USA \\
}
{\it 
$^2$William I. Fine Institute of Theoretical Physics,
University of Minnesota, Minneapolis, MN 55455, USA \\
}

\end{center}

\vspace{0.5cm}
\begin{abstract}
\noindent 
While LHC8 Higgs mass and sparticle search constraints favor a multi-TeV value of 
soft SUSY breaking terms, 
electroweak naturalness favors a superpotential higgsino mass $\mu\sim 100-200$ GeV: 
the mis-match results in an apparent Little Hierarchy characterized by $\mu\ll m_{\rm soft}$ (with $m_{\rm soft}\sim m_{3/2}$ in gravity-mediation).
It has been suggested that the Little Hierarchy arises from a mis-match 
between Peccei-Quinn (PQ) and hidden sector intermediate scales $v_{PQ}\ll m_{\rm hidden}$. 
We examine the Murayama-Suzuki-Yanagida (MSY) model of radiatively-driven
PQ symmetry breaking which not only generates a weak scale value of $\mu$ but 
also produces intermediate scale Majorana masses for right-hand neutrinos.
For this model, we show ranges of parameter  choices with multi-TeV values of $m_{3/2}$ 
which can easily generate values of $\mu\sim 100-200$ GeV so that 
the apparent Little Hierarchy suggested from data emerges quite naturally.
In such a scenario, dark matter would be comprised of an axion plus a 
higgsino-like WIMP admixture where the axion mass and higgsino masses are linked by the
value of the PQ scale.
The required light higgsinos should ultimately be detected at a 
linear $e^+e^-$ collider with $\sqrt{s}>2m({\rm higgsino})$.
\vspace*{0.8cm}

%\noindent PACS numbers: 12.60.Jv,14.80.Va,14.80.Ly

\end{abstract}

\end{titlepage}

\section{Introduction}

While the recent discovery of the Higgs boson with mass $m_h=125.5\pm 0.5$ GeV 
at the CERN LHC~\cite{atlas_h,cms_h} confirms the particle content of the Standard Model (SM), 
many physicists nonetheless expect new physics beyond the SM to yet emerge. 
This expectation arises theoretically from two fine-tuning problems that
afflict the SM: one in the electroweak sector arising from quadratically  divergent
contributions to the Higgs mass 
while the other arises in the QCD sector and is known as the strong CP problem~\cite{peccei}. 

The latter of these is solved elegantly by hypothesizing the existence of a global $U(1)_{PQ}$
symmetry~\cite{pq} valid at some high energy scale~\cite{ksvz,dfsz}, $v_{PQ}\sim f_a\sim 10^9-10^{16}$ GeV, where $v_{PQ}$ is the scale of the PQ symmetry breaking and $f_a$ is the axion decay constant.\footnote{It is model-dependent to determine the exact relation between $v_{PQ}$ and $f_a$. 
In most cases, $v_{PQ}\sim f_a$. 
We show the exact relation for MSY model (Ref.~\cite{msy}) in Sec.~\ref{sec:results}.}
Upon breaking of PQ symmetry, %~\cite
the axion field emerges as the associated massless Goldstone boson~\cite{ww}. 
The axion field acquires a mass and hence a potential due to QCD instanton effects.
In this case, then the offending CP-violating term 
\be
{\cal L}\ni \left(\bar{\theta}-\frac{a}{f_a}\right)G^{A\mu\nu}\tilde{G}^A_{\mu\nu}
\ee
can dynamically settle to tiny values. 
In the process, the universe is filled with a cold axion fluid --via the
mis-alignment mechanism-- which acts as cold dark matter (CDM)~\cite{axdm}.

The EW fine-tuning (or big hierarchy) problem is elegantly solved by 
introducing supersymmetry (SUSY) which guarantees cancellation of quadratic divergences~\cite{kaul}. 
The softly broken minimal supersymmetric SM (MSSM) then requires superpartners 
for all SM states which are expected to lie at or around the weak scale, 
since indeed some soft masses and the superpotential $\mu$ parameter contribute directly to the
Higgs, $W$ and $Z$ masses~\cite{primer,wss}. 
While indirect support for SUSY exists via gauge coupling unification and the measured values of
the top quark and Higgs boson mass, so far no superparticles have been seen at LHC. 
This latter situation is summarized by mass limits $m_{\tg}\gtrsim 1.3$ TeV (for $m_{\tg}\ll m_{\tq}$) and 
$m_{\tg}\gtrsim 1.8$ TeV (for $m_{\tg}\sim m_{\tq}$) in the context of simple models such as 
mSUGRA/CMSSM~\cite{atlas_s,cms_s}.
In models of gravity-mediated SUSY breaking, one expects SUSY to be broken in a hidden sector
so that the gravitino gains a mass $m_{3/2}\sim m_{\rm hidden}^2/M_P$ where $m_{\rm hidden}$ 
is some mass scale associated with the hidden sector and $M_P$ is the reduced Planck scale~\cite{nilles}. 
The effect of hidden sector SUSY breaking on 
the observable sector is to induce soft SUSY breaking terms of order $m_{3/2}$ 
in the Lagrangian so that the gravitino mass sets the scale for the sparticle masses~\cite{sw}. 
Based on recent LHC8 search limits, we thus expect 
$m({\rm sparticle})\sim m_{3/2} \gtrsim $ TeV which would then imply $m_{\rm hidden}\gtrsim 10^{11}$ GeV.

In contrast to the expectations for soft term masses given above, it is important to note that
the $W$, $Z$ and $h$ masses also depend on soft SUSY breaking terms and the superpotential
$\mu$ term via the shape of the (radiatively corrected) scalar potential which determines the Higgs field vevs
$v_u$ and $v_d$. For the $Z$ mass, we have
\be
\frac{m_Z^2}{2} = \frac{(m_{H_d}^2+\Sigma_d^d)-(m_{H_u}^2+\Sigma_u^u)\tan^2\beta}{(\tan^2\beta -1)}
-\mu^2\simeq -m_{H_u}^2-\mu^2
\label{eq:mzs}
\ee
where $m_{H_u}^2$ and $m_{H_d}^2$ are the {\it weak scale} soft SUSY breaking Higgs masses, 
$\mu$ is the {\it supersymmetric} higgsino mass term and $\Sigma_u^u$ and $\Sigma_d^d$ contain
an assortment of loop corrections to the scalar potential~\cite{rns}.
To avoid large, unnatural cancellations between $m_{H_u}^2$ and $\mu^2$ 
in obtaining the measured value of $m_Z$, one then expects that $|m_{H_u}^2|$ and $\mu^2$
are both $\sim m_Z^2$~\cite{rns,rnsltr,ccn,over,xt,seige}. 
The mis-match between LHC8 search limits and naturalness implies
a puzzling Little Hierarchy~\cite{bs} characterized by
\be
\mu\sim |m_{H_u}|\sim 100\ {\rm GeV}\ll m_{3/2}\sim 2-20\ {\rm TeV} .
\label{eq:LH}
\ee

The soft term $m_{H_u}^2$ is expected to be $\sim m_{3/2}^2$ at some high scale (usually taken to 
be $m_{\rm GUT}\simeq 2\times 10^{16}$ GeV). However, $m_{H_u}^2$ is driven radiatively 
through zero to negative values in the heralded radiative electroweak symmetry breaking (REWSB)
mechanism due to the large top-quark Yukawa coupling~\cite{rewsb}.
One simple way to accommodate naturalness is to accept that $m_{H_u}^2$ has been driven to
small rather than large negative values. Such a scenario has been dubbed
``radiatively-driven natural SUSY'' or RNS for short~\cite{rns,rnsltr}.

In addition to $m_{H_u}^2$, naturalness also expects that $\mu^2\sim m_Z^2$. 
However, since the $\mu$ parameter arises in the superpotential ({\it i.e.} it is 
supersymmetric and not SUSY breaking), naively one expects it to be of order the reduced 
Planck mass $M_P\simeq 2.4\times 10^{18}$ GeV.
This mis-match in expectations is known as the supersymmetric $\mu$ problem~\cite{kn,gm}.
Solutions to the $\mu$ problem first invoke some symmmetry to forbid the appearance
of $\mu$ in the superpotential. Next, the up- and down- Higgs multiplets are coupled to
new singlet fields either via renormalizable (NMSSM~\cite{nmssm}) or 
non-renormalizable (KN~\cite{kn} or GM~\cite{gm}) operators suppressed by powers of $M_P$.
Finally, one arranges for the singlets to gain suitable vevs so that an effective weak scale value of
$\mu$ is induced. 

In the Kim-Nilles solution~\cite{kn} to the $\mu$ problem, one introduces PQ charges for 
the Higgs fields $H_u$ and $H_d$ along with a PQ-charged field $\hat{X}$ coupled via
\be
\hat{f}_{KN}\ni \lambda_\mu \hat{X}^2 \hat{H}_u \hat{H}_d /M_P
\ee
which is in fact just the supersymmetrized DFSZ axion model which solves the strong CP
problem~\cite{chun}. 
The KN superpotential also includes the term
\be
\hat{f}_{KN}\ni \lambda_{PQ}\hat{Z}\left(\hat{X} \hat{Y}-v_{PQ}^2/2\right)
\ee
which causes the scalar components $\phi_X$ and $\phi_Y$ 
to gain vevs of order the PQ breaking scale 
$v_{PQ}/\sqrt{2}$ where $v_{PQ}=f_a/\sqrt{2}$
Then a $\mu$ term is induced with 
\be
\mu\sim \lambda_\mu f_a^2/M_P .
\ee
Originally Kim and Nilles had sought to relate the scales $f_a$ and $m_{\rm hidden}$. 
Instead, we see that the emerging Little Hierarchy characterized by $\mu\ll m_{3/2}$ 
may just be a consequence of a disparity between intermediate mass scales  
\be 
f_a\ll m_{\rm hidden} .
\ee
While it is sufficient phenomenologically to accommodate the PQ/hidden sector hierarchy by hand, 
it would be more satisfying to see such a hierarchy emerge naturally from a particle physics model.

A model which accomplishes such a goal has in fact been proposed some time ago by 
Murayama, Suzuki and Yanagida (MSY)~\cite{msy,gk}.
In the MSY model, the PQ scale $v_{PQ}$ emerges quite naturally in that PQ symmetry is radiatively 
broken as a consequence of SUSY breaking, much like the case where EWSB emerges as a consequence of
SUSY breaking. The question then is: does the MSY model (or other similar models) generate
a $\mu$ value comparable to $m_{3/2}$, or one that is comparable to $m_Z$ or $m_h$ as expected 
by naturalness? We will show in this paper
that the latter possibility emerges easily for generic model parameters, showing that 
values of $\mu$ comparable to $m_Z$ 
can be generated from TeV-scale values of $m_{3/2}$ (as seemingly required by LHC8 constraints).
Thus, the Little Hierarchy seems to lose some of its mystery, and one can reconcile naturalness
with the Higgs mass $m_h$ and LHC8 sparticle mass bounds.

To this end, in Sec.~\ref{sec:MSY} we review features of the MSY model which are relevant for 
our calculations. In Sec.~\ref{sec:results} we present our numerical results showing that natural
values of $\mu$ can be easily generated from multi-TeV values of $m_{3/2}$. Since the PQ scale
$v_{PQ}$ is related to $\mu$, then the Higgs mass, and better yet the higgsino masses if they are 
discovered, would provide an important clue as to the value of the axion mass.
An additional feature of the MSY model is that it generates simultaneously a third intermediate
mass scale-- the Majorana mass scale $M$ associated with the neutrino see-saw mechanism.

In such a model, we expect dark matter to be composed of a mixture of higgsino-like WIMPs
(but with non-negligible gaugino components) along with axions. The exact abundances of each
depend on details of the SUSY axion model~\cite{kimrev} (such as axino and saxion masses, PQ breaking scale
and saxion field strength) and computations for the SUSY DFSZ model have been presented previously in Ref.~\cite{dfsz2}.\footnote{In Ref.~\cite{dfsz2}, the effective theory was considered so that only axion superfield remains light among fields in PQ breaking sector ({\it e.g.} Kim-Nilles). In the MSY model, there is one light fermion and one complex scalar in addition to axion, saxion and axino. Although the decay processes of PQ particles are more complicated than those in Ref.~\cite{dfsz2}, the big picture is almost the same since all the couplings are still of order $\mu/f_a$.}
In Sec.~\ref{sec:conclude} we present our conclusions: mainly that the Little Hierarchy
Problem is no problem at all, but a {\it feature} to be expected in SUSY axion models which 
simultaneously address the gauge hierarchy problem, the strong CP problem and the SUSY $\mu$ problem. 

\section{MSY model of radiatively broken PQ symmetry}
\label{sec:MSY}

The MSY model assumes a MSSM superpotential of the form
\be
\hat{f}_{\rm MSSM}=\sum_{i,j=1,3}\left[
({\bf f}_u)_{ij}\epsilon_{ab}\hat{Q}^a_i\hat{H}_u^b\hat{U}^c_j +
({\bf f}_d)_{ij}\hat{Q}^a_i\hat{H}_{da}\hat{D}^c_j +
({\bf f}_e)_{ij}\hat{L}^a_i\hat{H}_{da}\hat{E}^c_j +
({\bf f}_\nu)_{ij}\epsilon_{ab}\hat{L}^a_i\hat{H}_u^b\hat{N}^c_j\right] .
\label{eq:Rcons}
\ee
where $\hat{N}^c$ is the SM gauge singlet field containing a right-hand neutrino.
The PQ charges are assumed to be $1/2$ and $-1$ for matter and Higgs fields, respectively. 
The MSSM superpotential is augmented by an additional set of terms containing 
new PQ charged fields $\hat{X}$ and $\hat{Y}$ with charges $-1$ and $+3$:
\be
\hat{f}'=\frac{1}{2}h_{ij}\hat{X}\hat{N}^c_i\hat{N}^c_j +\frac{f}{M_P}\hat{X}^3\hat{Y}
+\frac{g}{M_P}\hat{X}\hat{Y}\hat{H}_u\hat{H}_d .
\ee
For simplicity, $h_{ij}$ is taken as diagonal in generation space: $h_{ij}=h_i\delta_{ij}$ and we will also assume $h_1=h_2=h_3\equiv h$. 

The corresponding soft SUSY breaking terms are given by
\bea
V_{\rm soft}&=&m_X^2|\phi_X|^2+m_Y^2|\phi_Y|^2+m_{N_i^c}^2|\phi_{N_i^c}|^2 \nonumber\\
&&+ \left(\frac{1}{2}h_iA_i\phi_{N_i^c}^2\phi_X+\frac{f}{M_P}A_f\phi_X^3\phi_Y
+\frac{g}{M_P}A_g H_uH_d\phi_X\phi_Y+h.c.\right) .
\label{eq:Vsoft}
\eea

From these, one may compute the two-loop renormalization group equations (RGEs) by using the recipe in Ref.~\cite{2loop}. 
Neglecting neutrino Yukawa couplings, we find
\bea
\frac{dh_i}{dt}&=&\frac{h_i}{(4\pi)^2}\left(2|h_i|^2+\frac{1}{2}\sum_j|h_j|^2\right)-\frac{h_i}{(4\pi)^4}\left(2|h_i|^4+\sum_j|h_j|^4+|h_i|^2\sum_j|h_j|^2\right)\label{eq:RGEh}\\
\frac{dA_i}{dt}&=&\frac{2}{(4\pi)^2}\left(2|h_i|^2{A_i}+\frac{1}{2}\sum_j|h_j|^2{A_j}\right) \nonumber\\
&&-\frac{4}{(4\pi)^4}\left(2|h_i|^4A_i+\sum_j|h_j|^4A_j+\frac{1}{2}|h_i|^2A_i\sum_j|h_j|^2+\frac{1}{2}|h_i|^2\sum_j|h_j|^2A_j\right)\label{eq:RGEA}\\
\frac{dm_X^2}{dt}&=&\frac{1}{(4\pi)^2}\sum_i|h_i|^2\left(m_X^2+2m_{N_i^c}^2+|{A_i}|^2\right)\nonumber\\
&&-\frac{4}{(4\pi)^4}\sum_i |h_i|^4\left(m_P^2+2m_{N_i^c}^2+2|A_i|^2\right)\label{eq:RGEmP}\\
\frac{dm_Y^2}{dt}&=&0\label{eq:RGEmQ}\\
\frac{dm_{N_i^c}^2}{dt}&=&\frac{2|h_i|^2}{(4\pi)^2}\left(2m_{N_i^c}^2+m_X^2+|{A_i}|^2\right) -\frac{4|h_i|^4}{(4\pi)^4}\left(2m_{N_i^c}^2+m_X^2+2|A_i|^2\right)\nonumber\\
&&-\frac{|h_i|^2}{(4\pi)^4}(2m_{N_i^c}^2\sum_j|h_j|^2+2m_X^2\sum_j|h_j|^2+2\sum_jm_{N_j^c}^2|h_j|^2
\nonumber\\
&& +2A_i\sum_jA_j|h_j|^2+\sum_j|h_j|^2|A_j|^2+|A_i|^2\sum_j|h_j|^2)
\label{eq:RGEmN}
\eea
with $t=\ln(Q/M_P)$.
For simplicity, we will take all soft terms equal to $m_{3/2}$ or $m_{3/2}^2$ at 
$Q=M_P$ although this simplification need not apply.

One may then evolve the couplings and soft terms from $Q=M_P$ the reduced Planck scale
$M_P\simeq 2.4\times 10^{18}$ GeV down to the scale $Q\sim v_{PQ}$ of PQ symmetry breaking.
The essential feature here is that the soft mass $m_X^2$ gets driven radiatively
to negative values, resulting in the spontaneous breaking of PQ symmetry. 

The scalar potential consists of the terms $V=V_F+V_D+V_{\rm soft}$.
For now, we can ignore the Higgs field directions since these develop vevs at much lower
energy scales in radiatively-driven natural SUSY. Then the relevant part of the scalar
potential is just
\be
V_F\ni \frac{|f|^2}{M_P^2}|\phi_X^3|^2+\frac{9|f|^2}{M_P^2}|\phi_X^2\phi_Y|^2 .
\ee
Augmenting this with $V_{\rm soft}$, we minimize $V$ at a scale $Q=v_{PQ}$ 
to find the vevs of $\phi_X$ and $\phi_Y$ ($v_X$ and $v_Y$):
\bea
0&=& \frac{9|f|^2}{M_P^2}|v_X^2|^2v_Y +f^*\frac{A_f^*}{M_P}v_X^{*3}+m_Y^2v_Y 
\label{eq:minQ}\\
0&=& \frac{3|f|^2}{M_P^2}|v_X^2|^2v_X+\frac{18|f|^2}{M_P^2}|v_X|^2|v_Y|^2v_X
+3f^*\frac{A_f^*}{M_P}v_X^{*2}v_Y^*+m_X^2v_X .
\label{eq:minP}
\eea
The first of these may be solved for $v_Y$. Substituting into the second, we find a 
polynomial for $v_X$ which may be solved for numerically. 
The potential has two minima in the $v_X$ and $v_Y$ plane 
symmetrically located with respect to the origin. 
For practical purposes, 
we use the notation $v_X$=$|v_X|$ and $v_Y$=$|v_Y|$ in the rest of the paper.

At this point one may generate the Majorana neutrino mass scale 
\be
M_{N_i^c}=v_X \: h_i|_{Q=v_X}
\label{eq:Mnu}
\ee
and the SUSY $\mu$ term: 
\be
\mu =g\frac{v_Xv_Y}{M_P}\;.
\label{eq:mu}
\ee
Note that since the $\mu$ term depends on an arbitrary coupling $g$, one may obtain any desired
value of $\mu$ for particular $v_X$ and $v_Y$ vevs by suitably adjusting $g$. However, if the 
required values of $g$ are very different from unity, {\it i.e.} $g\gg1$ or $g\ll1$, we might need to introduce an additional physical scale 
to explain the $\mu$ term.
%large $\gtrsim 1$, then the theory likely becomes non-perturbative, while
%if tiny values of $g\ll 1$ are required then one might wonder about naturalness.

The QCD axion field $a$ is now the corresponding Goldstone boson of the broken PQ symmetry
and is a combination of the phases of the $\phi_X$ and $\phi_Y$ fields. 
Along with the axion, one gains a corresponding saxion $s$ and axino $\ta$ with masses
$\sim m_{3/2}$ but with superweak couplings suppressed by $1/v_{PQ}$. 
In addition, one obtains an orthogonal combination of 
a super-weakly coupled singlet field $\phi_s$ plus a singlino $\ts$
also with masses $\sim m_{3/2}$.

\section{Numerical results}
\label{sec:results}

In this section, we report on results of our numerical solution of the 
coupled RGEs~(\ref{eq:RGEh})-(\ref{eq:RGEmN}) and subsequent determination 
of the PQ scalar vevs via Eq's.~(\ref{eq:minQ}) and (\ref{eq:minP}).
The vevs $v_X$ and $v_Y$ allow us to determine the values of the Majorana neutrino 
intermediate scale $M$, Eq.~(\ref{eq:Mnu}), and the SUSY $\mu$ parameter, Eq.~(\ref{eq:mu}).
\begin{figure}[tbp]
\includegraphics[height=0.4\textheight]{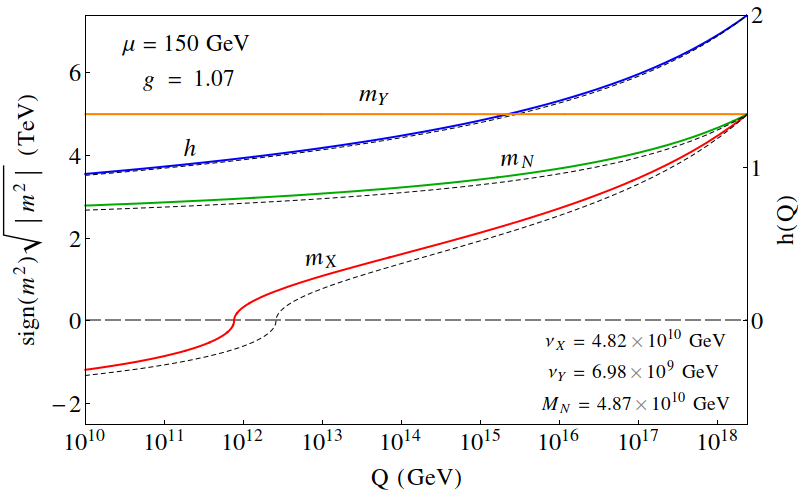}
\caption{Plot of the running values of various soft terms and couplings versus $Q$
for $h=2$. 
We take a common value of SUSY breaking parameters, {\it i.e.} $m_X=m_Y=m_{N_i^c}=A_i=5$ TeV. 
Black dashed lines show RG evolution without the 2-loop corrections.
\label{fig:run2}}
\end{figure}
\begin{figure}[tbp]
\includegraphics[height=0.4\textheight]{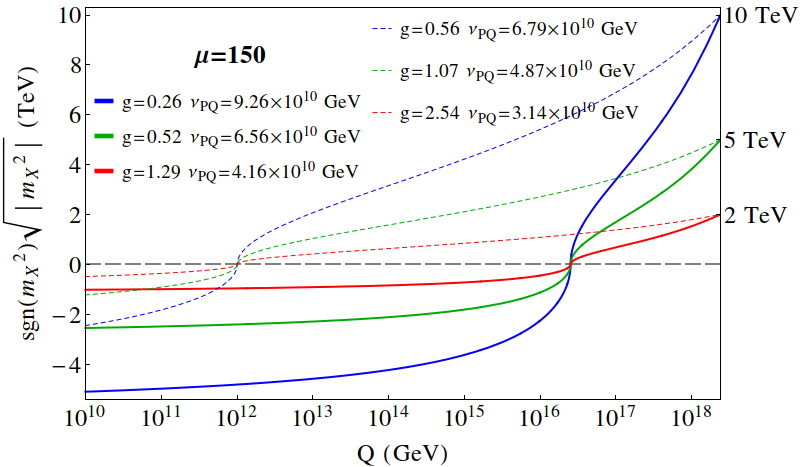}
\caption{Plot of the running values of $m_X^2$ versus $Q$ for various 
values of $m_{3/2}$ and $h=2$ (dashed) and $h=4$ (solid).
\label{fig:run_m32}}
\end{figure}

In Fig.~\ref{fig:run2} we show a case of the coupled RG evolution of PQ soft terms and couplings 
versus renormalization scale $Q$ starting from the reduced Planck mass $M_P$ down to 
the scale of PQ breaking.\footnote{Although Fig.~1 and Fig.~2 show the evolution of parameters 
down to $10^{10}$ GeV, we find solutions for $g$ and $M_N$ at $Q=v_{PQ}$ 
for each set of parameters.} In the figure, we adopt a PQ-neutrino coupling value $h_i=2$ and assume
universal SUSY breaking parameters set equal to $m_{3/2}$ at $M_P$ with value $m_{3/2}=5$ TeV.
While $m_Y^2$ remains constant, $m_{N_i^c}^2$ is suppressed by RG running. Meanwhile, the value
of $m_X^2$ is pushed from an initial value of 5 TeV down through zero to negative values
so that PQ symmetry is radiatively broken. Solving the scalar potential 
minimization conditions (with canonical choices $f=1$ and $A_f=-m_{3/2}$)
implies values of $v_X=4.82\times 10^{10}$ GeV and $v_Y=6.98\times 10^{9}$ GeV.
The PQ scale $v_{PQ}=\sqrt{v_X^2+v_Y^2}=4.87\times10^{10}$ GeV so indeed an intermediate scale PQ breaking
is generated.
In this case, the axion decay constant is $f_a=\sqrt{v_X^2+9v_Y^2}=5.26\times 10^{10}$ GeV.\footnote{The axion model is of DFSZ type so the axion interaction is determined by 
$f_a/N_{\rm DW}$ where $N_{\rm DW}=6$.}
Furthermore, a right-hand (RH) Majorana neutrino scale is generated to be 
$M_{N^c_i}=4.78\times 10^{10}$ GeV. Finally, a SUSY $\mu$ term is also generated. In this case, a value of
$g=1.07$ in the PQ superpotential $\hat{f}'$ allows for a value of $\mu=150$ GeV which is the
expected region from naturalness.

In Fig.~\ref{fig:run_m32}, we show the RG running of the critical soft breaking mass $m_X^2$
versus energy scale $Q$ for several initial values of $m_X=2$, 5 and 10 TeV. We also take values of $h_i= 2$
(dashed curves) and 4 (solid curves) at $M_P$. In the case of the dashed curves with $h_i=2$, we see
that for each case of $m_X$, the value of $m_X^2$ gets driven negative at exactly the same value of 
$Q$ so that PQ symmetry is broken in each case. By solving the minimization conditions, we
are able to generate a value of $\mu =150$ GeV by adopting values of $g=2.54$, 1.07 and 0.56 
for $m_X(Q=M_P)=2$, 5 and 10 TeV respectively. Thus, indeed a multi-TeV value of SUSY breaking
soft parameters can generate a value of $\mu\sim m_Z$ as required by naturalness and 
resulting in a Little Hierarchy. If instead we take $h_i=4$, then values of
$g=1.29$, $0.52$ and $0.26$ are required to generate $\mu =150$ GeV for 
$m_X(M_P)=2$, 5 and 10 TeV.

\begin{figure}[tbp]
\includegraphics[height=0.4\textheight]{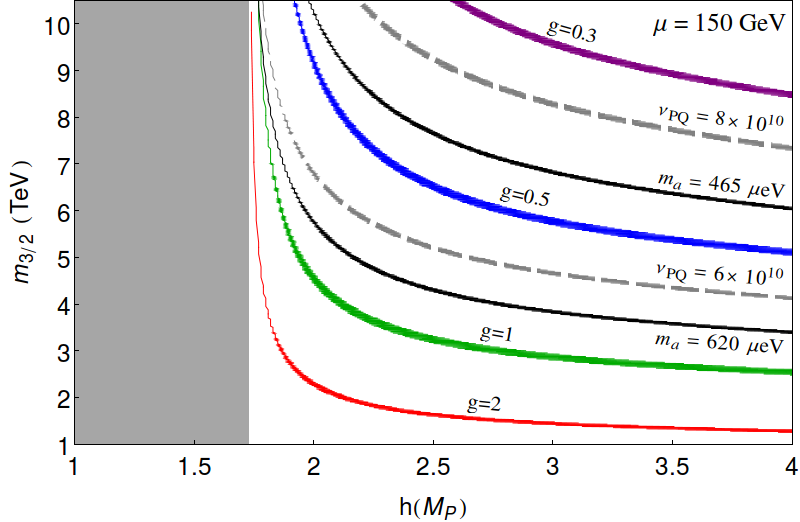}
\caption{Values of $g$ which are needed to generate $\mu =150$ GeV 
in the $h$ vs. $m_{3/2}$ plane. 
Dashed gray lines show contours of constant $v_{PQ}$ and black lines 
show contours of constant axion mass, $m_a$.
\label{fig:h_m32}}
\end{figure}
In Fig.~\ref{fig:h_m32}, we plot contours of the value of $g$ which is required to 
generate a $\mu$ parameter of 150 GeV in the $h(M_P)$ versus $m_{3/2}$ plane. 
The first point to note is that if $h_i$ is too small, then $m_X^2$ will not get driven
negative. In the case in which parameters run down to $10^{10}$ GeV ($\sim v_{PQ}$), this region occurs for $h_i\lesssim 1.73$ and is shaded gray.
%For higher values of $h_i$, namely $h_i\gtrsim6$, two-loop RGE correction terms become dominant at high energy scales and as a result $m_X^2$ is driven to negative values with much steeper slope. This evolution of $m_X^2$ changes the behaviour of the contour lines in Fig.~\ref{fig:h_m32} at $h\sim 6$ where the slopes tend to get larger values. For $h_i\gtrsim8.3$, there is no physical solution for PQ breaking since $m_X^2$ at Q=$10^{10}$ GeV takes infinitely large values. With only one-loop RGEs, for higher values of $h_i$, we would have almost horizontal contour lines: 
%then PQ symmetry would always be broken and a value of $\mu\sim m_Z$ could be generated.
Typically, in the $h(M_P)$ versus $m_{3/2}$ plane, 
large values of $g>1$ are required for rather low values of $m_{3/2}\lesssim 2.5$ TeV. For much 
higher values of $m_{3/2}\sim 5$ TeV, then typically $g\sim 0.5$ is required to generate the 
Little Hierarchy. Values of $g\sim 0.2-0.3$ can generate $\mu=150$ GeV for $m_{3/2}$ in the 10 TeV range. 
We also show contours of $v_{PQ}=6\times 10^{10}$ GeV, 
$v_{PQ}=8\times 10^{10}$ GeV (dashed gray lines), 
$m_a=465$ $\mu$eV and $m_a=620$ $\mu$eV (black lines) on the same plane. 
In the region above the $g=2$ line, $f_a$ and $v_{PQ}$ can take 
a range of values such as $3.7\times 10^{10} \lesssim f_a \lesssim 1.1\times 10^{11}$ GeV 
and $3.4\times 10^{10} \lesssim v_{PQ} \lesssim 9.4\times 10^{10}$ GeV. 
%We should note that 
%$f_a$ and $v_{PQ}$ can exist within the same limits
%for $m_{3/2}\gtrsim 10$ TeV and small enough $h_i$.

\begin{figure}
\hfill
\subfigure[]
%\subfigure[Plot of value of $\mu$ vs. variation in trilinear soft term $A_f$ 
%for three values of $m_{3/2}$ and for $f=g=1$ and $h=2$.]
{\includegraphics[width=8.4cm]{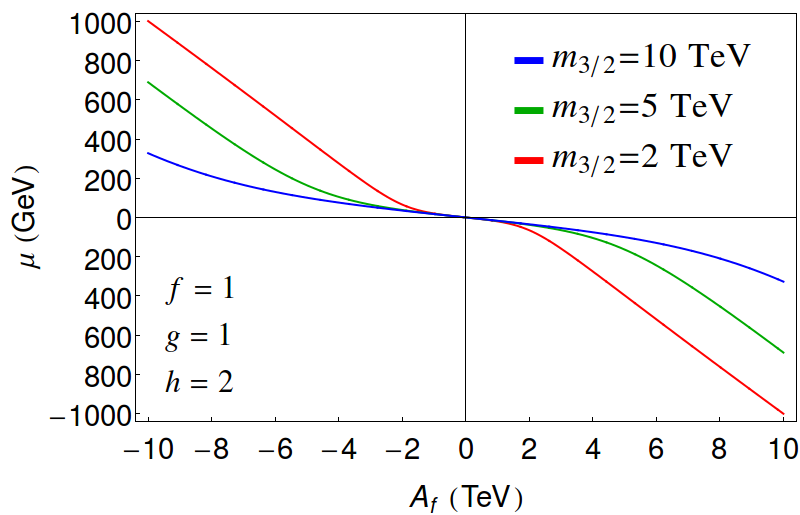}}
\hfill
\subfigure[]
%\subfigure[Plot of value of $\mu$ vs. variation in $f$ 
%for three values of $m_{3/2}$ and for $g=1$, $h=2$ and $A_f=-m_{3/2}$.]
{\includegraphics[width=8.2cm]{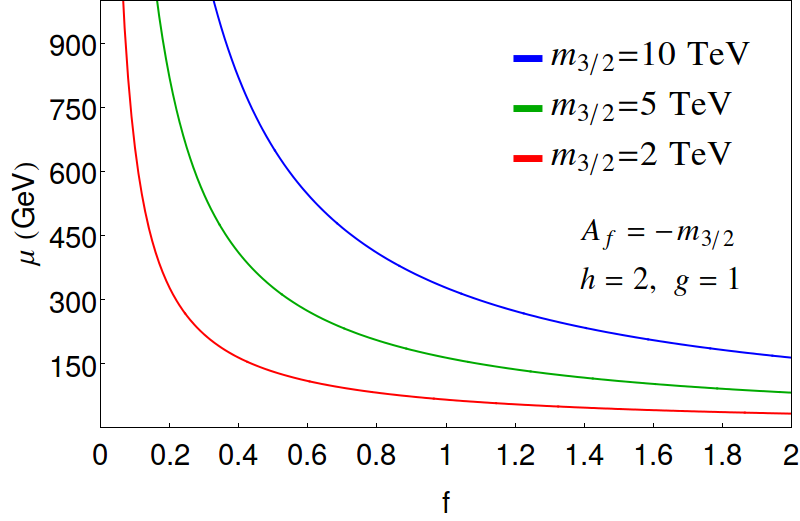}}
\hfill
\caption{Plot of value of $\mu$ for three values of $m_{3/2}$ 
vs.  ({\it a}) variation  
in $A_f$ for $f=g=1$ and $h=2$ and ({\it b}) 
variation in $f$ for $g=1$, $h=2$ and $A_f=-m_{3/2}$.}
\label{fig:mu}
\end{figure}

While the solution for $m_X^2$ is independent of $f$ and $A_f$, the vevs $v_X$ and $v_Y$
do depend on these quantities, and hence so does $\mu$. In our previous plots, we have taken
a canonical choice $f=1$ and $A_f=-m_{3/2}$. By choosing $A_f< 0$ we get vevs with the same sign which generate positive $\mu$ values, this choice has no other effect on any results.\footnote{We can also obtain positive $\mu$ by taking $g<0$ for $A_f>0$.}
In Fig.~\ref{fig:mu}{\it a}, we show the 
value of $\mu$ which is generated versus $A_f$ by taking $m_X=m_Y=m_{3/2}$ for 
$m_{3/2}=2$, 5 and 10 TeV with $f=g=1$ and $h=2$. With $A_f=0$, we generate
$v_Y=0$ so that $\mu =0$ while for $A_f\sim m_{3/2}$, then we generate
natural values for $\mu\sim 100-200$ GeV. For very large 
$|A_f|\gg m_{3/2}$, then unnaturally large values of $\mu$ develop 
for the lower range of $m_{3/2}\sim 1-2$ TeV.
In Fig.~\ref{fig:mu}{\it b}, we show the variation of $\mu$ versus $f$.
In this case, we see that very small values of $f$ result in large
$v_X$ and $v_Y$ and hence large $\mu $ values. 
%As $f$ approaches towards $0$, the vevs increase so that $\mu$ becomes large.
For $f\sim 1$, then natural values of $\mu\sim 100-200$ GeV can develop.

A phenomenological aspect of the MSY model has been investigated 
by Martin~\cite{spm1}.
Since two PQ fields $\hat{X}$ and $\hat{Y}$ have been hypothesized, then one combination
gives the usual axion-axino-saxion supermultiplet while the other gives a 
super-weakly coupled singlet-singlino combination $(\phi_s ,\ts )$ with masses $\sim m_{3/2}$. 
While normally one would not expect such super-weakly coupled states to give rise to collider effects, 
in this case the singlino $\ts$ could be the LSP. Then each NLSP produced 
via sparticle production followed by cascade decays in collider experiments would ultimately decay to
the singlino. These delayed NLSP decays could give rise to sparticle production events 
with displaced vertices which might be easily seen in LHC detectors.

\section{Some related models}
\label{sec:models}

In the previous section, we have seen that, starting with multi-TeV values of 
gravitino mass (as required by LHC constraints for models of gravity mediation)
one can easily generate values of $\mu\sim m_Z$ as required by electroweak naturalness.
In this case, a Little Hierarchy emerges quite naturally from radiatively-driven PQ symmetry breaking.
While our results are illustrated in the MSY model of radiative PQ symmetry breaking, the 
overall phenomena may be more general. Here we comment on two related models.

A very similar model is written down by Choi, Chun and Kim (CCK)~\cite{cck}. In the CCK model, 
the PQ part of the superpotential is given by
\be
\hat{f}_{CCK}=\frac{1}{2}h_{ij}\hat{X}\hat{N}^c_i\hat{N}^c_j +\frac{f}{M_P}\hat{X}^3\hat{Y}
+\frac{g}{M_P}\hat{X}^2\hat{H}_u\hat{H}_d .
\label{eq:cck}
\ee
While the PQ charge assignments will differ from the MSY case, 
this model also exhibits radiative PQ symmetry breaking for sufficiently large values of $h_i$. 
Thus, the resulting $\mu$ term is similar to the MSY case.
%The group CCK also remark that it is easy to write down
%models of radiatve PQ breaking in the supersymmetrized KSVZ model with intermediate scale heavy quark
%superfields. In this case, however, there is no associated solution to the SUSY mu problem.

Martin has also written down similar models but with a different mechanism for PQ 
breaking~\cite{primer,rpc,spm2}.
In this case, the superpotential is given by
\be
\hat{f}_{SPM}\ni \frac{g_1}{M_P}\hat{X}^2\hat{H}_u\hat{H}_d +\frac{g_2}{M_P}\hat{X}^2\hat{Y}^2 .
\label{eq:spm}
\ee
Martin notes that the field directions $\hat{X}$ and $\hat{Y}$ give rise to nearly flat directions 
in the scalar potential. In such a case, then Planck-suppressed hard SUSY breaking quartic operators
are expected to occur and can contribute to the scalar potential. Then
one can achieve intermediate scale PQ breaking even without soft mass terms being driven to 
negative values~\cite{spm2}.
It is also possible to break PQ symmetry in the MSY model by the large quartic coupling, 
{\it i.e.} large $|A_f|$ in Eq.~(\ref{eq:Vsoft}).
In this case, however, the PQ scale is rather large, {\it i.e.} $v_{PQ}\sim m_{\rm hidden}$, 
and thus we need much smaller $g$ to generate a  natural value of $\mu\sim m_Z$.
Models with more than two PQ fields are of course also possible.

\section{Conclusions} 
\label{sec:conclude}

In this paper, we have explored the case where the gauge hierarchy problem is solved via supersymmetry
while the strong CP problem is solved by the introduction of PQ symmetry and its concommitant axion.
In such models, three intermediate scales are present: the hidden sector mass scale $m_{\rm hidden}$, 
the Majorana neutrinos scale $M_N$ and the PQ scale $v_{PQ}$. We have explored consequences of the 
MSY SUSY axion model which is able to generate the neutrino and PQ scales as a consequence of
radiative PQ symmetry breaking triggered by hidden sector SUSY breaking. In fact, in string theory
the first expectation is that the PQ scale $f_a\sim M_{\rm GUT}-M_P$~\cite{witten}. In the MSY model instead it 
naturally emerges at a phenomenologically more viable intermediate scale $\sim 10^{10}-10^{12}$ GeV.

While LHC sparticle search limits plus the rather high value of the Higgs mass $m_h\sim 125.5$ GeV
seem to indicate a sparticle mass scale $m_{3/2}$ in the multi-TeV range, electroweak
naturalness requires the weak scale soft term $|m_{H_u}|$ and the $\mu$ parameter to be of order
$m_Z$. While $m_{H_u}^2$ can be driven to small negative values via radiative electroweak symmetry
breaking, the MSY model provides a similar mechanism to produce a value of $\mu\sim 100-200$ GeV
via radiative PQ breaking. In this case, the Little Hierarchy characterized by 
$\mu\ll m_{3/2}$ emerges quite naturally and is in fact associated with the intermediate
scale hierarchy $f_a\ll m_{\rm hidden}$. In this class of models, one expects dark matter to
be composed of an axion plus higgsino-like WIMP admixture, and detection of both 
should ultimately be expected~\cite{bbm}.
While sparticles may or may not be detected at LHC~\cite{lhc}, 
the expected light higgsinos should definitely be detected at a linear $e^+e^-$ collider~\cite{ilc}
provided that $\sqrt{s}>2m({\rm higgsino})$. In such a case, the measured value of $\mu\sim f_a^2/M_P$ 
will be related to the axion mass $m_a\sim 620\mu{\rm eV}\left(10^{10}\ {\rm GeV}/(f_a/N_{DW})\right)$.

\section*{Acknowledgments}

We thank Xerxes Tata and Azar Mustafayev for comments on the manuscript.
This work was supported in part by the US Department of Energy, Office of High Energy Physics.

%
%%%%%%%%%%%%%%%%%%%%%%%%%%%%%%%%%%%%%%%%%%%%%%%%%%%%%%

%
\end{document}